\documentclass[twocolumn,tighten, times, trackchanges]{aastex61}

\usepackage{color}
\usepackage{soul}

\shorttitle{massive compact galaxies at $1.0 < z < 2.0$}
\shortauthors{Gu et al.}

\begin{document}

\title{Structures, stellar population properties, AGN fractions, and environments of massive compact galaxies at $1 < z < 2$ in 3D--{\it HST}/CANDELS}

\correspondingauthor{Qirong Yuan and Guanwen Fang}
\email{yuanqirong@njnu.edu.cn, wen@mail.ustc.edu.cn}

\author{Yizhou Gu}
\affil{School of Physics Science and Technology, Nanjing Normal University,
Nanjing  210023, China;
yuanqirong@njnu.edu.cn}

\author{Guanwen Fang}
\altaffiliation{Guanwen Fang and Yizhou Gu contributed equally to this work}
\affil{Institute for Astronomy and History of Science and Technology, Dali University, Dali 671003, China; wen@mail.ustc.edu.cn
}

\author{Qirong Yuan}
\affil{School of Physics Science and Technology, Nanjing Normal University,
Nanjing  210023, China;
yuanqirong@njnu.edu.cn}

\author{Shiying Lu}
\affil{School of Physics Science and Technology, Nanjing Normal University,
Nanjing  210023, China;
yuanqirong@njnu.edu.cn}







\begin{abstract}
We present a study on structures and physical properties of massive ($M_* >10^{10} M_{\sun} $) compact galaxies at $1.0<z<2.0$ in five 3D--{\it HST}/CANDELS fields.
Compared with the extended star-forming galaxies (eSFGs), compact star-forming galaxies (cSFGs) are found to have the lower level of star formation, and mainly distribute in the quiescent region of the {\it UVJ} diagram.
The distributions of dust attenuation and S{\'e}rsic index support that the progenitors of cQGs are cSFGs, and cSFGs are at a transitional phase between eSFGs and cQGs.
The prevalence of X-ray selected AGNs  ($\sim 28\%$) is confirmed in the cSFGs at $1<z<2$ which indicates that the violent gas-rich processes such as merger and disk instability could drive the structure to be more compact, and trigger both star formation and black hole growth in the central regions.
Our results support the ``two-step'' scenario that the cSFGs at $1<z<2$ are the intermediate population after compaction but before a quick quenching.
Our analysis of parametric and nonparametric morphologies shows that cQGs (eQGs) are more concentrated and have less substructures than cSFGs (eSFGs), and quenching and compactness should be associated with each other.
The cSFGs at $1.5<z<2$ ($1<z<1.5$) prefer to be in higher (lower) density environment, similar as cQGs (eSFGs).
It suggests that merger or strong interaction might be the main driving mechanism of compaction at higher redshifts, whereas the disk instability of individual galaxies might play a more important role on the formation of cSFGs at lower redshifts.
\end{abstract}




\section{Introduction} \label{sec:intro}
How galaxies terminate their star formation activities is one of the key questions in extragalactic astronomy.
It is reported that a small number of quiescent galaxies at $z \sim 3-4$ are identified by photometric and/or spectroscopic data \citep{Guo+12, Gobat+12, Muzzin+13, Straatman+14, Glazebrook+17}. Individual quiescent galaxies begin to emerge when the age of the universe was roughly 1.5 Gyr, corrsponding to z = 4, indicating that a portion of star-forming galaxies (SFGs) is even being quenched at very early time.
Since the SDSS-based works reveal the bimodality in color distribution \citep{Strateva+01, Baldry+04}, the galaxy bimodality has been confirmed to exist up to $z \sim 2$ at least (\citealt{Ilbert+10, Brammer+11, Muzzin+13}). It means that quiescent galaxies have been an important population since $z \sim 2$.

Compared to local early-type galaxies, a distinguishing feature of quiescent galaxy population at high redshifts is their smaller sizes of light profiles.
On average, quiescent galaxies in early epoch are 3 to 5 times more compact than their local counterparts (e.g., \citealt{Daddi+05,Trujillo+07, vD+08, vD+10, Whitaker+12, Barro+13, vdW+14}).
It has been confirmed that the number density of compact quiescent galaxies (cQGs) progressively increases with cosmic time at $2<z<3$, and declines with cosmic time at $z<1$ \citep{Barro+13, Cassata+13, vD+15}. cQGs, which are referred to as `red nuggets',  are more common at high redshifts, whereas they are extremely rare in the nearby universe (\citealt{Taylor+10, Buitrago+18}).
The number density of cQGs reaches a peak around $z\sim 1.5$ and decreases at lower redshifts regardless of the compactness criteria adopted \citep{Charbonnier+17, Lu+19}.

Typical blue galaxies are disc-dominated or irregular galaxies with extended structures and considerable efficiencies of star formation, and they are referred to as extended star-forming galaxies (eSFGs).
Interestingly,  some `blue nuggets'  (i.e., compact star-forming galaxies, cSFGs)  are found at high redshifts (\citealt{Wuyts+11b, Barro+13}), which possess the similar compact structures but are still forming stars. Massive cSFGs are found to have a plateau in the number density at $2 < z < 3$ and a continuous drop from $z \sim 2$ to 1, and rare cSFGs are found at lower redshifts $0.5 < z < 1$ \citep{Barro+13, Barro+14a, vD+15}.

However, the origin and fate of these cSFGs are not clearly understood yet. Naturally, the cSFGs have been considered as a bridge to build the evolutionary connection between eSFGs and cQGs. Besides the quenching models, another physical process, compaction, is indispensable to complete the evolutionary track.
The progenitors of the cSFGs at $2 < z < 3$ are expected to be the eSFGs undergoing a compation phase at $z > 3$ or at this same epoch \citep{Barro+14a, Barro+14b, Wellons+15, Wellons+16}. Massive cSFGs can also be formed from
the low-mass compact galaxies undergoing the rapid growth of stellar mass \citep{Williams+15}.
The cSFGs at high redshifts could not maintain their star-forming status for long, and would rapidly convert into cQGs due to some quenching mechanisms such as AGN and stellar feedbacks \citep{Barro+13,Tadaki+15}. Then, these cQGs are proposed to evolve into extended quiescent galaxies (eQGs) and into the local giant ellipticals eventually via the later size growth by merger potentially (\citealt{Naab+07, Newman+12, vD+15, delaRosa+16}).
The evolutionary sequence via compaction and quenching is called `early track' (see \citealt{Barro+13}), which is expected to happen at early time ($z>2$).
Besides, the other track (`late track') is that eQGs can be formed from eSFGs directly, without the need of compaction phase, by some gas-poor quenching mechanisms, such as secular processes \citep{KK+04} or strangulation \citep{Peng+15}.

Statistics of physical properties of compact galaxies at higher redshifts, such as number density, size, structural parameters, velocity dispersion and so on, suggests that the cSFGs at $z = 2 - 3$ are the direct progenitors of the cQGs (e.g., \citealt{Barro+13, Barro+14a, Barro+14b, Nelson+14, Tadaki+15, Fang+15, vD+15, Straatman+15, Barro+16b, Barro+17b}).
\cite{Barro+16a} capture one cSFG in the act of rapid quenching or in a transition stage at $z \sim 1.7$.
Recently, some evidences are found that the formation of cQGs could also occur at $z < 1$ (\citealt{Charbonnier+17, NC+18, NC+19, Damjanov+19}).
The compact galaxies at $1 < z <2$ is of great important to investigate the quenching and compaction processes in galaxy evolution, which connects the higher redshifts ($2<z<3$) where the cQGs have increasing densities with cosmic time and the lower redshifts ($z<1$) where the numbers of cQGs and cSFGs remarkably decrease.
The period from $z \sim 2$ to 1 stand astride the inflection point of the number density of cQGs, and it is possible to observe the competition in the number density of cQG at $1<z<2$ between the quenching process forming new cQGs and later size growth which results in the transformation of cQGs into eQGs.

The physical properties of massive cSFGs at $2 < z < 3$ have  been discussed \citep{Barro+14a, Fang+15}.
In order to supply the evolutionary scenario of the compact galaxies at $1<z<2$, we construct a large sample of compact galaxies with $M_*>10^{10}M_{\sun}$ over this redshift range in five 3D--{\it HST}/CANDELS fields (\citealt{Grogin+11, Koekemoer+11,Skelton+14, Momcheva+16}).
According to the thresholds of specific star-forming rate (sSFR $\equiv {\rm SFR}/M_*$) and
galaxy compactness $({\rm\Sigma_{1.5}}\equiv M_*/r_e^{1.5}$),
all massive galaxies at $1<z<2$ are divided into 4 subsamples, which are termed as eSFGs, cSFGs, cQGs, and eQGs, respectively.
To examine whether the early track could occur at $1<z<2$, we investigate number densities, stellar population, structural parameters, AGN fractions, and localized environments for the eSFGs, cSFGs, cQGs, and eQGs.
Physical mechanisms which may dominate the formation of eQGs and cQGs from $z = 2$ to 1 will be also addressed.

The paper is outlined as follows. In section \ref{sec:data}, we introduce the 3D--{\it HST} and CANDELS data, the details of the galaxy properties, and sample selection. In section \ref{sec:prop}, we present physical properties of the eSFGs, cSFGs, cQGs and eQGs at $1<z<2$.  Possible quenching mechanisms concerning the compaction and quenching scenario are discussed in section \ref{sec:disc}. Finally, we summarize our conclusions in section \ref{sec:sum}.

Throughout our paper, we assume the cosmological parameters as following: $H_0=70\,{\rm km~s}^{-1}\,{\rm Mpc}^{-1}$, $\Omega_m=0.30$, $\Omega_{\Lambda}=0.70$.

\section{Data and sample selection} \label{sec:data}
\subsection{Redshifts, Rest-Frame Colors, and Stellar Masses}
The CANDELS programs acquire the {\it HST} observations covering $\sim$ 900 arcmin$^2$ in five fields:  AEGIS, COSMOS, GOODS-N, GOODS-S, and UDS  (\citealt{Grogin+11, Koekemoer+11}).
The 3D--{\it HST} Treasury Program\footnote{\url{https://3dhst.research.yale.edu/Home.html}} provides a large amount of the data products, refering to photometry \citep{Skelton+14} and grism spectra \citep{Momcheva+16}, together with the derived parameters, such as structural parameters \citep{vdW+14} and star formation rates (SFRs; \citealt{Whitaker+14}).

\cite{Momcheva+16} provide a ``best'' redshift catalog by merging their grism-based results with the catalog from \citealt{Skelton+14}. The ``best'' redshifts are constructed by the following order: spectroscopic redshifts, grism redshifts, and photometric redshifts. Spectroscopic redshifts are collected from the ground-based telescopes (seeing \citealt{Skelton+14} for detail). The spectral energy distributions (SEDs) ranging from 0.3$-$8.0 $\mu m$ are used to derive photometric redshifts with the EAZY code (\citep{Skelton+14}).
The uncertainty of photometric redshifts reaches $\Delta z /(1+z) \approx 0.02$ on average. The estimate of grism redshift is determined by the combined `spectrum+photometry' fitting, using a modified EAZY code. For most galaxies, their grism redshifts are of extremely high accuracy with $\Delta z /(1+z) \approx 0.003$ \citep{Momcheva+16}.


Once the redshifts are well constrained as the "best" redshifts, the stellar population parameters, rest-frame colors  , and SFRs are derived by \citealt{Momcheva+16}. The rest-frame colors can be derived from the filter response function and the best-fit template for individual sources with the EAZY code. Stellar population parameters, including stellar and star formation timescale, are estimated with the FAST code (\citealt{Kriek+09}), by assuming exponentially declining star formation histories, solar metallicity, \cite{Calzetti+00} dust extinction law and \cite{BC03} stellar population synthesis models with a \cite{Chabrier+03} initial mass function (IMF).
SFRs are determined by the contributions of UV emission from massive stars and the IR emission that re-radiated by dust,  which derived by the following luminosity conversion (\citealt{Kennicutt+98, Bell+05}):
\begin{equation}
{\rm SFR_{UV+IR}}[M_{\sun} \cdot {\rm yr^{-1}}]=1.09\times10^{-10}(L_{\rm IR}+2.2L_{\rm UV})/L_{\sun},
\end{equation}
where $L_{\rm UV}$ represents the rest-frame 1216-3000\AA \ luminosity which is  determined by using the 2800\AA \ rest-frame luminosity times a factor of 1.5, {\bf $L_{\rm UV} = 1.5 \nu L_{\rm \nu,2800}$}.
This factor is attributed to the shape of UV spectrum for a 100 Myr old population with a constant SFR.
Infrared luminosity, $L_{\rm IR}$, means the integrated 8-1000 $\mu m$ luminosity, which is converted from the observed $Spitzer$/MIPS 24 $\mu m$ flux density using a single luminosity-independent template (\citealt{Wuyts+08}). The $Spitzer$/MIPS 24 $\mu m$ photometric catalogue is provided by \citealt{Whitaker+14}.

For the galaxies undetected at 24 $\mu m$,  the correcting UV-based SFR for dust attenuation is adopted.
The dust-corrected $\rm SFR_{UV, corr}$ is derived from the rest-frame near-ultraviolet luminosities ($\rm \sim 2800\AA$) after correcting for dust attenuation ($\rm A_V$) from the SED fitting results.
Assuming the \cite{Calzetti+00} dust attenuation curve, the dust-corrected $\rm SFR_{UV, corr}$ are derived by
\begin{equation}
{\rm SFR_{UV, corr}}[M_{\sun} \cdot {\rm yr^{-1}}]={\rm SFR}_{\rm UV}\times10^{0.4\times 1.8 \times \rm A_V},
\end{equation}
where ${\rm SFR_{UV}}=3.6 \times 10^{-10} \times  L_{2800}/L_{\odot}$ assuming a \cite{Chabrier+03} IMF, following \cite{Wuyts+11a}. $\rm A_V$ is the SED-based optical attenuation provided by the FAST, and the factor of 1.8 corresponds to the \cite{Calzetti+00} attenuation curve at 2800\AA.
\citet{Fang+18} make a comparison between these two SFR estimates, and reveal small but significant systematic differences between $\rm SFR_{UV,corr}$ and $\rm SFR_{UV+IR}$.


\subsection{Structural Parameters}
Galaxies can be modeled with a single S\'{e}rsic profile \citep{Sersic+68}.
Based on public {\it HST}/WFC3 F125W and F160W imaging from CANDELS, the structural parameters, such as S\'{e}rsic index $n$, half-light radius $r_{hl}$, and axis ratio $b/a$, have been well measured with GALFIT (\citealt{Peng+02}) by \cite{vdW+14}. In this work, the  morphologies at  rest-frame 5500 \AA\ are traced by $J$-band (F125W) images at $1.0<z<1.5$ and $H$-band (F160W) images at $1.5<z<2.0$.
We quantify the galaxy size using the circularized effective radius, $r_{e} = r_{hl} \times \sqrt{b/a}$, rather than using $r_{hl}$ directly, where $r_{hl}$ is the half-light radius along major axis.

The nonparametric measurements are also introduced. We take the measurements using the program Morpheus developed by \cite{Abraham+07} on both $J$-band (F125W) and $H$-band (F160W) images.
The definitions of the model-independent quantities, including concentration index ($C$), Gini coefficient ($G$), and second-order moment of the brightest 20\% light ($M_{20}$), are described as follow (also see the review in \citealt{Conselice+14}).

Concentration index ($C$) describes the concentration of the surface brightness distribution in the center of a galaxy. It is defined as the ratio between two integral fluxes within the inner isophotal radius, $0.3R$, and within the outer isophotal radius, $R$,
\begin{equation}
C = \frac{F_{0.3R}}{F_{R}},
\end{equation}
where $R$ is the radius for a enclosed area by galaxy isophote at 2$\sigma$ level above the sky background \citep{Abraham+94}. Hubble sequence from the irregulars to the ellipticals is also the sequence that $C$ increases.

Gini coefficient ($G$) is a statistical measure of inequality degree of light distributions within a segmentation map of each galaxy \citep{Lotz+04}. It is defined as
\begin{equation}
    G = \frac{ \sum_{i}^{N}(2i - N- 1)\left| F_i \right|}{\left|\bar{F} \right|N(N-1)},
\end{equation}
where $F_i$ is the sorted flux in each pixel, $\bar{F}$ is a mean flux, and $N$ is the total number of pixels assigned to image segmentation of a galaxy. It ranges from 0 to 1, with 0 representing perfect equality for all pixels and 1 representing complete inequality as all fluxes assembling in a pixel.
$G$ is strongly correlated with the concentration degree of galaxy brightness which might be relative to several different locations, while $C$ only relative to the galactic center. For example, the galaxies with some bright components in the outskirts may result in a high $G$ but a relative low $C$. In other words, $G$ is independent of the spatial distribution of surface brightness.

The second order moment of the 20\% brightest pixels ($M_{20}$) is associated with sub-structures \citep{Lotz+04}.
It is defined as
\begin{equation}
  M_{20} = \log(\frac{\sum_{i} M_i}{M_{\rm tot}}) , {\rm with} \sum_i F_i < 0.2F_{\rm tot},
\end{equation}
where $M_i = F_i[(x_i - x_c)^2+(y_i - y_c)^2]$ and $M_{\rm tot} = \sum_{i=1}^{N}{M_i} $ for the brightest 20\% pixels in a galaxy.  $M_{20}$ increases as the presence of bars, spiral arms, or multiple bright cores in a galaxies. Typical $M_{20}$ value for late-type galaxies is $\sim$ -1.5, and that for early-type galaxies is about -2 \citep{Lotz+04}.

\begin{figure*}
\centering
\includegraphics[scale=1.0]{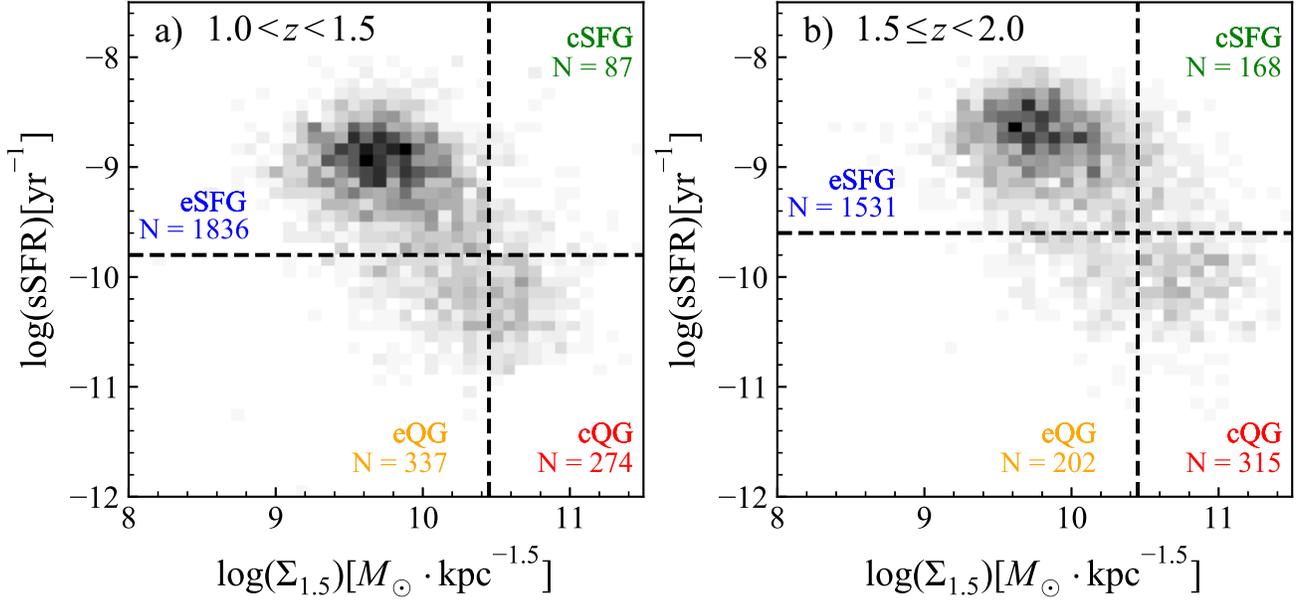}
\caption{The sSFR$-\Sigma_{1.5}$ diagram for massive galaxies at $1.0 < z < 2.0$.
The distribution of galaxies is denoted by the grey-scale map.
The vertical dashed line is our selection criterion for massive compact galaxies which is defined as $\Sigma_{1.5} = 10^{10.45} M_{\sun} \cdot {\rm kpc^{-1.5}}$, whereas the horizontal dashed line is the criterion for massive quiescent galaxies which is defined as sSFR equal to $10^{-9.8}$ and $ 10^{-9.6} {\rm yr^{-1}}$, respectively. The two thresholds separate our sample into four subsamples of eSFGs, cSFGs, cQGs and eQGs, which are shown in blue, green, red, and orange, respectively.
}
\label{fig01}
\end{figure*}

\subsection{Sample Selection}
In this study, we focus on the physical properties of massive compact galaxies at $1 < z < 2$ in all five 3D--{\it HST}/CANDELS fields. Massive extended galaxies are also involved for comparison. In the first place, only the objects with flag is selected {\tt use\_phot} = 1 (see \citealt{Skelton+14}). It means that the object (1) is not too faint and not a star, (2) is not contaminated by a bright source, (3) is well exposed both in the F125W and F160W, (4) has a signal-to-noise ratio (S/N) $> 3$ in F160W images, and (5) has “noncatastrophic” photometric redshift and stellar population fits.
After that, the galaxies in our sample are selected by applying the following criteria:
\begin{enumerate}
\item the galaxies brighter than 25 mag in $H$ band;
\item the galaxies with $M_* \geq 10^{10}M_\odot$ at $1 < z < 2$;
\item the galaxies with GALFIT quality flag= 0 (good fit) or flag= 1 (suspicious fit).
\end{enumerate}
The luminous type I AGNs are expected to be a potential concern, which would be severely contaminated by AGN emission. However, these luminous  AGNs ($\log L_X > 44$) are expected to be few ($<7\%$) based on the result of Galaxy+AGN modeling in GOODS-S \citep{Hsu+14}. Our selection of GALFIT quality flag exclude $\sim 6$ per cent of sources due to the bad quality flag = 2 or 3, whose the morphologies are hardly described by one single S\'ersic profile. This selection can help us discard some point-like sources which might be the luminous AGNs.
And the discarded galaxies do not occupy a special place of the redshift distribution and the ${\rm sSFR}–M_*$ or {\it UVJ} diagrams.
Noticed that host galaxies dominate the UV-to-NIR SEDs for most (90\%) X-ray AGNs, the measurements of the color, stellar mass, and SFR, so as morphology, should be reliable in five CANDELS fields (e.g., \citealt{Luo+10, Kocevski+17, Yang+17}). Thus, the discarded galaxies due to bad quality and the AGN contamination would not bias the main results.

The magnitude cut, $H_{\rm F160W} < 25$, guarantees the uncertainty $\Delta z/(1+z) \approx 0.02$ for some galaxies with photometric redshift alone \citep{Skelton+14, Bezanson+16}.
Moreover, the magnitude limit is set for the completeness of stellar mass down to $10^{10} M_\odot$ up to $z = 2$ and reliability of the structural measurements. Following the empirical method in \cite{Pozzetti+10}, the mass completeness limits as the magnitude cut $H_{\rm F160W} < 25$ are $10^{9.44} M_\odot$ at $1<z<1.5$ and $10^{9.83} M_\odot$ at $1.5<z<2$.
Almost all the galaxies (4734/4750) in our sample are $H_{\rm F160W} <$ 24.5, whose uncertainty of morphological parameters $r_{hl}$ and $n$ can be limited to less than 10\% \citep{vdW+12}. The 16 remaining galaxies with $H_{\rm F160W} > $ 24.5 take up quite a small number of our sample. All these fainter galaxies have the good fit flags =0. None of them are large galaxies with large S\'ersic indices, whose uncertainty can exceed 20\% at fainter magnitudes $H_{\rm F160W} > $ 24.5  \citep{vdW+12}. As a result, a mass-complete sample of 4750 massive galaxies are constructed at $1 < z < 2$. According to the sources of their ``best'' redshifts, only about 8\% of the galaxies in our sample are found to have spectroscopic redshifts, 51\% have grism redshifts, and 41\% have photometric redshifts.

Throughout the paper, the degree of galaxy compactness is quantified as the ratio of stellar mass to the 1.5 power of size, which is defined as $\Sigma_{1.5}\equiv M_*/r_e^{1.5}$ according to the size$-$mass relation for quiescent galaxies (\citealt{Shen+03, Newman+12, Barro+13}). Following \cite{Barro+14a}, we adopt the limit of compactness $\Sigma_{1.5} = 10^{10.45} M_{\sun} \cdot {\rm kpc}^{-1.5}$ to separate the compact from extended. The inverse of sSFR represents a timescale for the formation of stellar population with a constant SFR. Hence, sSFR is a relatively direct quantity to trace the degree of quiescence. To avoid ``cosmic downsizing'' at the high mass end (e.g. \citealt{Cowie+96}), we adopt the median sSFR of galaxies with $10^{10}<M_*<10^{10.5}M_\sun$ as the normalization of star formation main sequence (MS). Then we define the quiescent region with 0.8 dex below the MS. Considering the offset of median sSFR between two redshift bins is $\sim 0.2$ dex, we adopt the evolving thresholds in sSFR $10^{-9.8}$ and $10^{-9.6} \rm yr^{-1}$, respectively. 

According to the two quantities $\Sigma_{1.5}$ and sSFR, the sample is divided into the four subpopulations, namely eSFGs, cSFG, cQGs, and eQGs.
Figure \ref{fig01} shows both the definition for compact galaxies and the sSFR upper limit for quiescent galaxies.
The distribution of our massive galaxy sample in the $\Sigma_{1.5}-$sSFR plane is represented as the grey-scale map.
Among 4750 galaxies with $M_* > 10^{10} M_\odot$ and $1.0 < z < 2.0$, 539 ($\sim 11\%$) galaxies are compact and quiescent. This sample also contains 255 cSFGs, 589 eQGs, and 3367 eSFGs.
From $z \sim 1.75$ to $z \sim 1.25$, the comoving number densities of cSFGs decease from $(1.19 \pm 0.08) \times 10^{-4} {\rm Mpc}^{-3}$ to $(0.73 \pm 0.09) \times 10^{-4} {\rm Mpc}^{-3}$, whereas the comoving number densities of cQGs don't significantly evolving from $(2.23 \pm 0.13) \times 10^{-4} {\rm Mpc}^{-3}$ to $(2.30 \pm 0.14) \times 10^{-4} {\rm Mpc}^{-3}$. The  The net flow of compact galaxies is towards quenching preferentially, since the continuous formation of new cSFGs is slower than the continuous formation of cQGs via quenching. To sum cSFGs and cQGs up, the total number densities of compact galaxies drop about $\sim 0.4 \times 10^{-4} {\rm Mpc}^{-3}$ with time, which implies that some compact galaxies transform into extended.
The comparisons of structure and star formation properties between the eSFG, cSFG, and cQG populations would shed some more light on the physics processes concerning structural compaction and star formation quenching, respectively.

\begin{figure*}
\centering
\includegraphics[scale=1.0]{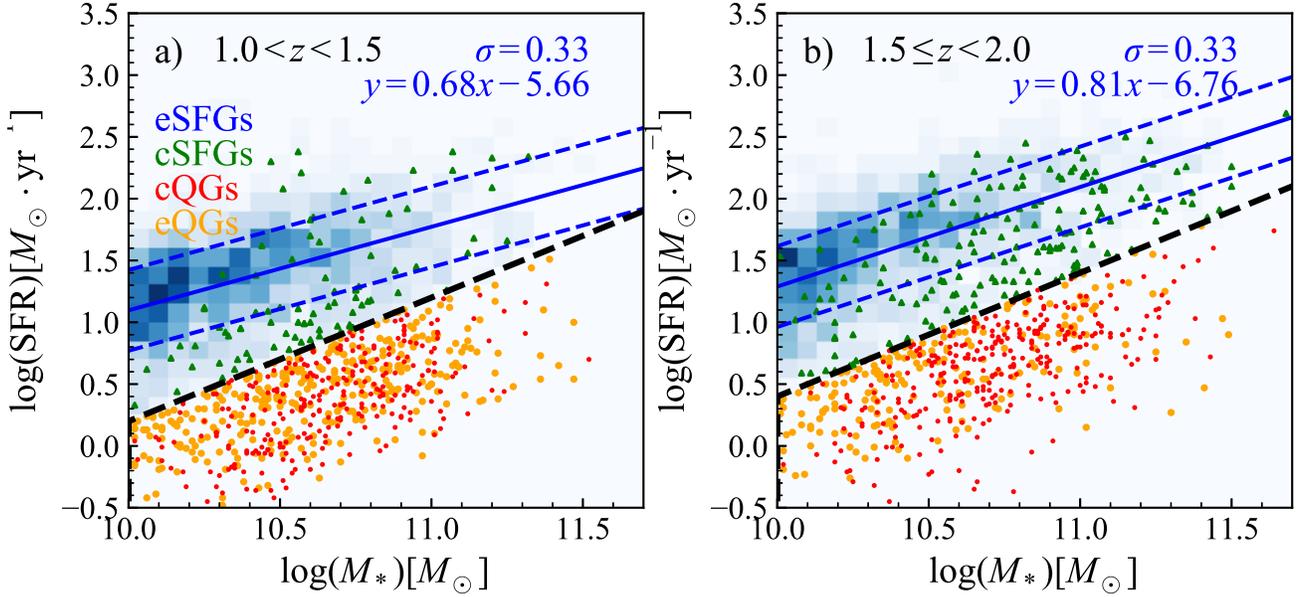}
\caption{SFR as a function of stellar mass for galaxies at $1.0 < z < 2.0$. The black dashed line is our selection criterion for massive quiescent galaxies which defines as sSFR equal to $10^{-9.8}$ and $ 10^{-9.6} {\rm yr^{-1}}$, respectively. Our sample of massive quiescent galaxies is shown as red (compact) and orange (extended) dots.
Boxed blue scales signify eSFGs, and green triangles signify cSFGs.
Blue solid lines are the best linear fitting for eSFGs with the blue dashed lines representing 1$\sigma$ dispersions.
}
\label{fig02}
\end{figure*}

\section{PHYSICAL PROPERTIES OF massive compact galaxies at $1.0 < z< 2.0$}
\label{sec:prop}
In this section, we focus on presenting a series of physical properties for the eSFGs, cSFGs, cQGs, and eQGs at $1 < z < 2$. Their star-forming statuses are explored with the SFR$-M_*$ and {\it UVJ} color diagrams. The properties of the stellar populations such as stellar mass, dust attenuation, and stellar age will also be presented. We also explore the galaxy morphologies from the perspective of the parametric measurements and our nonparametric measurements.

\subsection{The SFR$-M_*$ Relation}
For the star-forming galaxies within a wide range of redshift, there is a tight relation, known as the main sequence, between star formation rate and stellar mass \citep{Elbaz+07, Whitaker+14, Speagle+14}.
In Figure \ref{fig02}, we show the SFR as a function of stellar mass for eSFGs (boxed blue scales), cSFGs (green triangles), cQGs (red dots) and eQGs (orange dots) at $1.0 < z < 2.0$.  And we also perform a linear fitting of MS for eSFGs as a reference, which is shown as blue solid lines and dashed lines (1$\sigma$ dispersions).

We find our subsample of cSFGs contains only a few galaxies ($< 10\%$) with SFRs 1$\sigma$ over the MS for eSFGs. It is well consistent with previous works at $2<z<3$ where few cSFGs is found above the MS \citep{Barro+14a, Fang+15}. If the cSFGs are formed via merging or disk instability, they are expected to have experienced a gas-rich phase, accompanied by the trigger of starburst and/or AGN \citep{Dekel+14, Zolotov+15}.
It implies that the high level of star-formation activity cannot last too long since the cSFGs with high SFRs are less observed. After the compact phase, these might be a starburst phase with a very short timescale, and then cSFGs get into a quenching phase subsequently.

An appreciable quantity of cSFGs is the cSFG fraction that locate 1$\sigma$ below the MS of eSFGs.  A certain portion of eSFGs is found to have begun the quenching of star formation activities during their compaction processes.
The fraction of cSFGs below the MS increases from  45\% to 57\% as redshift decreases from $z \sim 1.75$ to $z\sim1.25$. What's more, \cite{Barro+14a} find that $\sim 30\%$ of cSFGs at $2< z < 3$ are 1 $\sigma$  below the MS. It is clear that the fraction of these cSFGs decrease with redshift, which is a strong evidence that some of these cSFGs are translating into cQGs.
It is reported in \citet{Fang+15} that cSFGs also have slightly lower SFRs than eSFGs at $2< z < 3$ . It supports that the overall star-forming status for cSFGs are turning down with the decrease of redshift, which implies that the cSFGs at the lower redshift are more likely to be quenched.
It indicates that this evolutionary status in the cSFG phase does not reach the dynamic equilibrium --- namely, the newly formed cSFGs are less than the quenched cSFGs at the same epoch.

\subsection{The {\it UVJ} Diagrams}
\begin{figure*}
\centering
\includegraphics[scale=1.0]{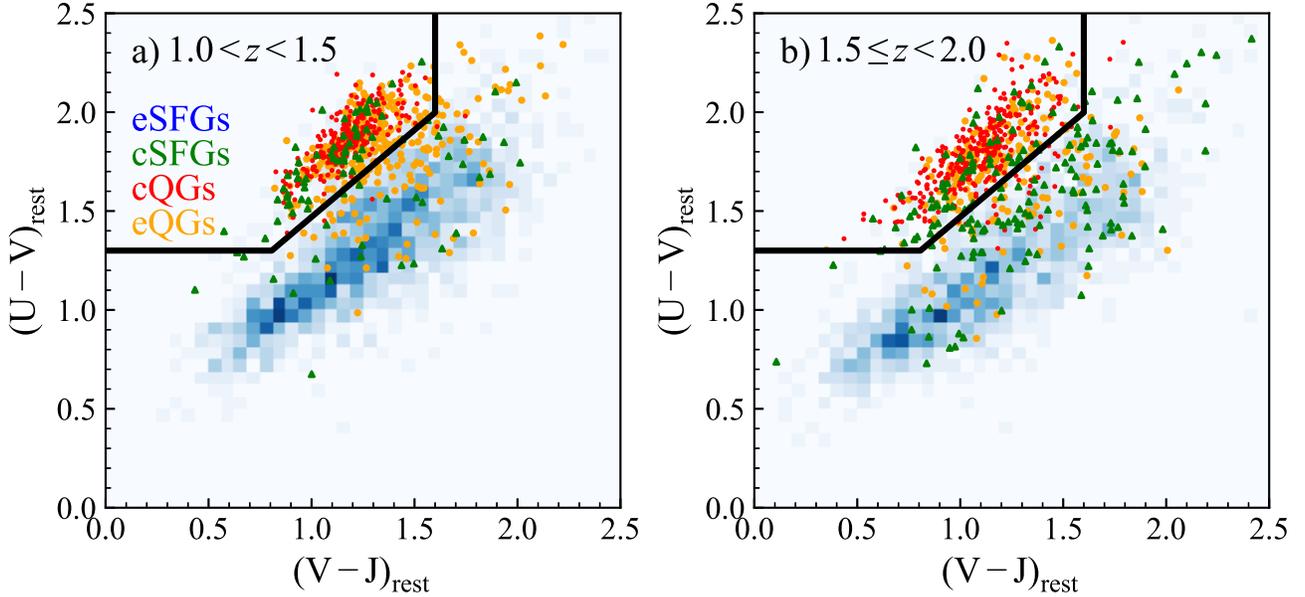}
\caption{Rest-frame {\it UVJ} diagram for four galaxy populations. The solid dividing line follows \citealt{Williams+09}. The colors and symbols are the same as those in Figure \ref{fig02}.\label{fig03}}
\end{figure*}

The {\it UVJ} diagram is a widely accepted diagnostic tool for separating the star-forming and quiescent populations (e.g., \citealt{Williams+09, Straatman+16, Fang+18}). We examine the star-forming status by the {\it UVJ} diagram, as shown in Figure \ref{fig03}. The solid dividing lines follow \cite{Williams+09}. The quiescent wedge is defined by $(U-V)>0.88(V-J)+0.49$, $(U-V)>1.3$, and $V-J<1.6$, whereas the star-forming region occupies the rest.

It is found that most quiescent galaxies (both cQGs and eQGs) occupy the quiescent region, whereas  eSFGs dominate in the star-forming region. Further, it shows a trend that the increasing faction of cSFGs from 38\% to 62\% are found in the {\it UVJ}-quiescent region with cosmic time.
We confirm that the cSFGs in the {\it UVJ}-quiescent region are almost corresponding to the cSFGs below the MS for eSFGs indeed. It means that the cSFGs occupying the quiescent region are found to be self-consistent with the galaxies with relatively lower sSFRs.
Hence, we conclude that the {\it UVJ} results are in accordance with the results in the SFR$-M_*$ plane. And a larger percentage of cSFGs trend to move into the quiescent region. It suggests that cSFGs represent a bridge between eSFGs and quiescent galaxies.

\subsection{Stellar Population Properties}
\begin{figure*}
\centering
\includegraphics[scale=0.6]{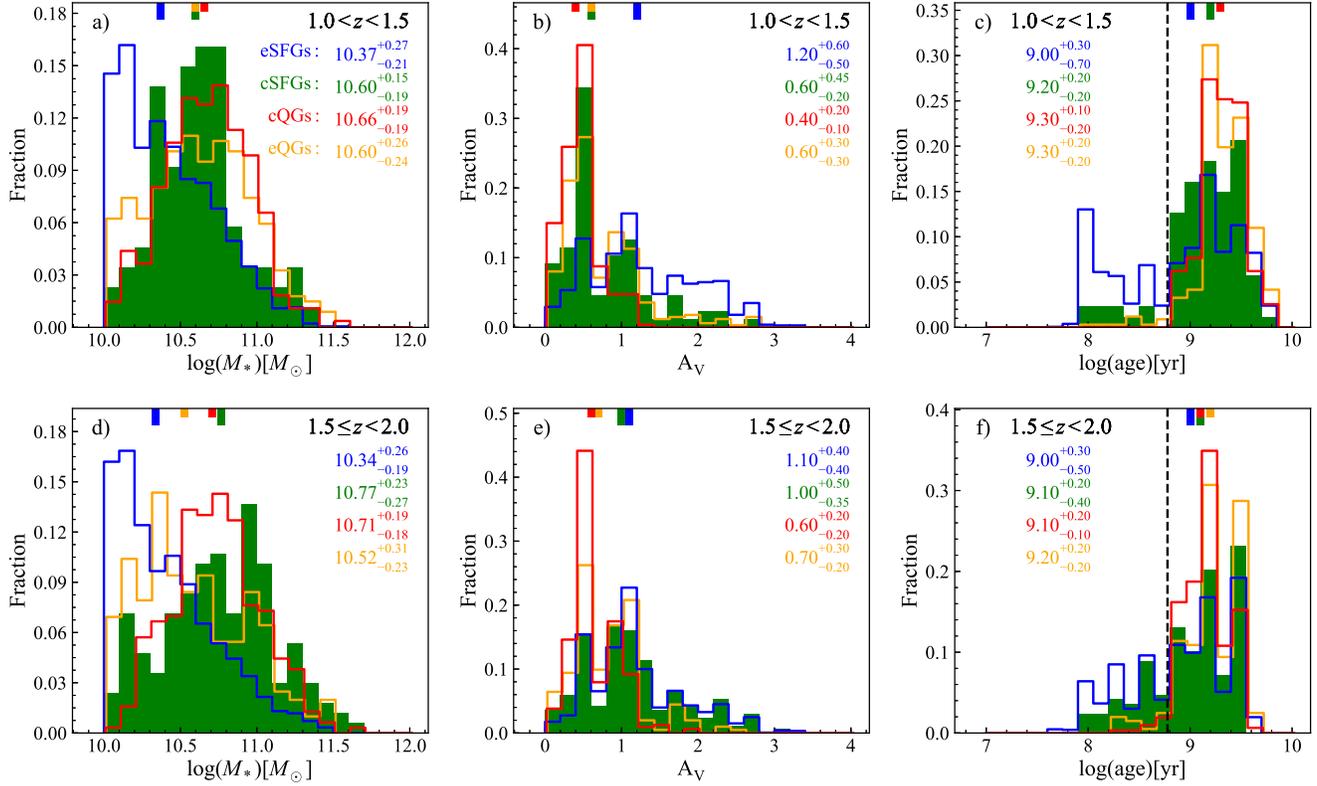}
\caption{Distributions of stellar mass, dust-extinction, stellar age,
for eSFGs, cSFGs, cQG, and eQGs, denoted by  blue, green, red, and orange histograms, respectively. The median values with uncertainty spanning the 25th to 75th percentiles of each distribution are shown in the corresponding colors. The dash lines in the right panels is the separating lines located at 0.6 Gyr, which define the galaxies with young stellar ages. 
}
\label{fig04}
\end{figure*}

Figure \ref{fig04} shows the distributions of  stellar population properties including stellar mass, dust attenuation, and stellar age. 
The median stellar mass of eSFGs is the lowest compared with other three  galaxy populations. And cSFGs possess the similar stellar mass with cQGs. If galaxies obey the two-step quenching scenarios (e.g. \citealt{Fang+13, Barro+17a, Wang+18}), eSFGs would become compact first and then quench into cQGs shortly after.
It is not surprised that cSFGs are significantly more massive than eSFGs.
Galaxies can accumulate a substantial stellar mass during the compaction phase due to the persistence of star formation.
During the subsequent quenching process, cQGs do not show a significant increase in stellar mass. It can be explained by the short quenching timescale so that galaxies are incapable of accumulating much stellar mass from star-forming phase to quiescent status.
The stellar mass distribution of eQGs at $1<z<2$ can be interpreted by the complicated origin of eQG population. The eQGs can be formed via both early track and late track (\citealt{Barro+13}). We will discuss the formation of eQGs in Section \ref{sec:eQGs}.

Panels (b) and (e) show that eSFGs have the heaviest and cQGs have the fewest extinctions, and cSFGs are in the middle. It is interesting to find that this result is inconsistent with the result of massive compact galaxies at $2 < z < 3$ (\citealt{Barro+14a, Fang+15}), which found that cSFGs possess the highest level of dust extinction. It can be explained by the different epochs. At $z > 2$, star formation activities are fierce so as the extinctions also increase (e.g, \citealt{Fang+15}). At $z<2$, the reduction of overall gas content and star formation level leads to the extinction of cSFGs in between. 
It is possible that dust extinction decreases in a sequence from eSFG to cSFG to cQG due to the overall continuing gas consumption with redshift, which supports that cSFGs are also at a transitional phase between eSFGs and cQGs at $1<z<2$.


Panels (c) and (f) exhibit the distributions of mean stellar ages for four populations in two redshift bins. It is believed that galaxies would experience a violent episode of star formation that move above the main sequence and also quickly build up compact stellar cores (e.g., \citealt{Toft+17}, \citealt{Gomez-Guijarro+19}). The age bimodility can be only found in the eSFGs at $1<z<1.5$. Assuming the galaxies with mean stellar age younger than 0.6 Gyr is ongoing recent starburst, we are witnessing their compaction phases traced by recent starburst activities.  Among these galaxies with younger stellar age, almost all occur in eSFGs at $1<z<1.5$, whereas cSFGs account for only a small component (around $1\%$). It hints that most cSFGs at $1<z<1.5$ finish the recent starbursts in their compaction phases.
At $1.5<z<2$, however, a larger percentage ($\sim7\%$) of these galaxies with younger stellar age can be found in cSFGs. It indicates that there are more cSFGs in the compaction progress via recent starburst activities at $1.5<z<2$. 
The properties of stellar populations support the two-step quenching scenarios. In the compaction phase, galaxies still retain a certain degree of star formation. The primary accumulation of the stellar mass is nearly completed after the compaction phase. Panel (c) shows that the distribution of stellar ages for the cSFGs at $1<z<1.5$ lies between those of eSFGs and cQGs. The increasing trend of the stellar age distributions from eSFGs to cSFGs to cQGs agrees well with the late track. Panel (f) shows that the median value of stellar ages for cSFGs are comparative with cQGs. Just like at $2<z<3$, the cSFGs at $1<z<2$ are also short-lived so that they would rapidly quench into cQGs without remarkable mass assembly.

\subsection{Galaxy Morphologies and Structures}

\subsubsection{Parametric Morphology}

\begin{figure*}
\centering
\includegraphics[scale=0.6]{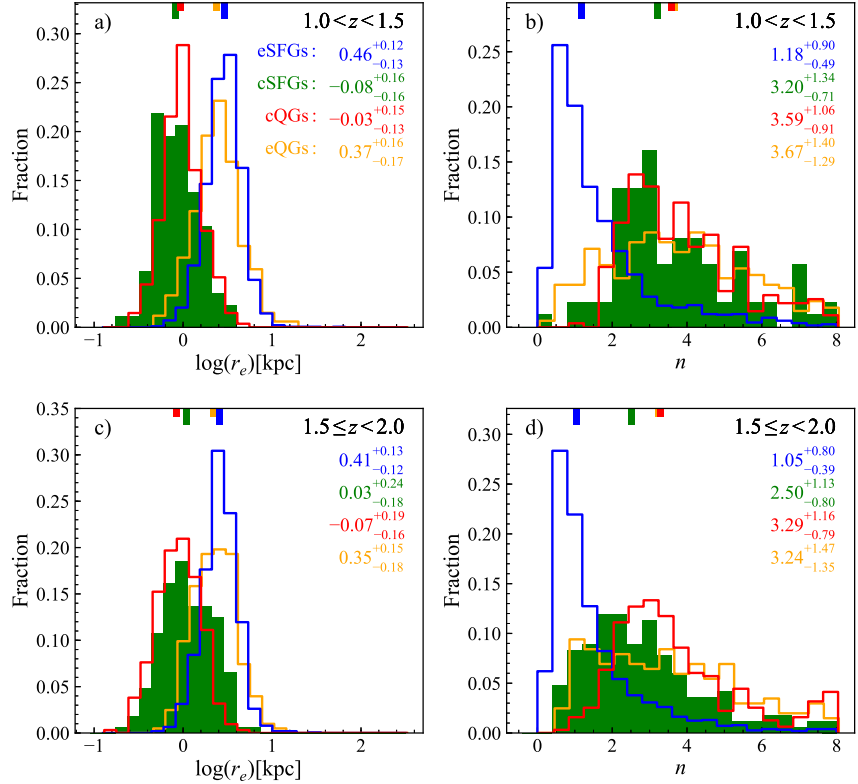}
\caption{Distributions of the circularized effective radius and S\'{e}rsic index for eSFGs, cSFGs, cQG, and eQGs, denoted by  blue, green, red, and orange histograms, respectively. The median values with uncertainty spanning the 25th to 75th percentiles of each distribution are shown in the corresponding colors.
}
\label{fig05}
\end{figure*}

The two dimensional profiles of galaxies can be modelled by the S\'{e}rsic model, and their structural parameters such as circularized effective radii and S\'{e}rsic indices can be derived.
Figure \ref{fig05} shows the distributions of these structural parameters for eSFGs, cSFGs, cQGs and eQGs.
The median size of eSFGs is similar to that of eQGs. And the median size of cSFGs is also similar to that of cQGs, whereas the median size of eSFGs is significantly larger than compact galaxies.
The eSFGs at $1<z<2$ show the disc-dominated morphologies with $n < 2.5$, and the cQGs are bulge-dominated with $n >2.5$.
For cSFGs, the median value of  S\'{e}rsic index falls in between those of eSFGs and cQGs. In the compaction phase, stellar mass tends to be redistributed at smaller radius due to some dissipative processes, such as disk instability and merge \citep{Dekel+14, Zolotov+15}. The median of S\'{e}rsic index ($n$) for cQGs is larger than that for cSFGs, indicating that the buildup of bulge component is accompanied with the quenching into cQGs.
Compared with eSFGs, eQGs show a more scattered distribution of S\'{e}rsic index.
Some eQGs possess the evidence of flattened discs with $n < 2.5$ (\citealt{Bruce+12, Barro+13}). For these galaxies, cQGs can grow their disk and become extended.
Both minor and major mergers may trigger the growth of outskirts and result in the rebuild of disk component (\citealt{Naab+09, Hopkins+09a}).

\subsubsection{Nonparametric Morphology}\label{sec:CGM}
The nonparametric diagnostics of galaxy structure provide a model-independent analysis on galaxy morphology.
The nonparametric structural measurements, including concentration index (C), Gini coefficient ($G$) and the second-order moment of the brightest 20\% light ($M_{20}$) , are presented in this section.

\begin{figure*}
\centering
\includegraphics[scale=1.0]{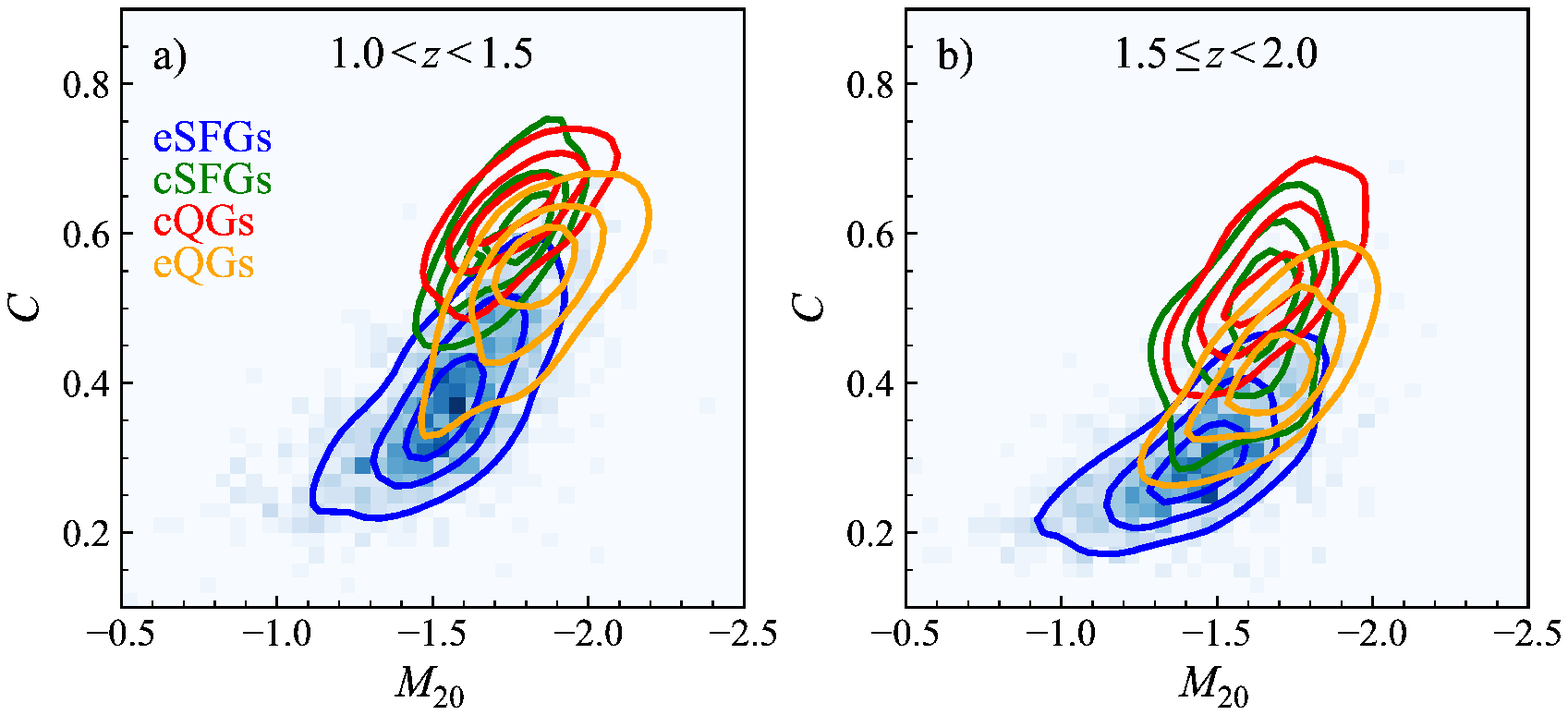}
\includegraphics[scale=1.0]{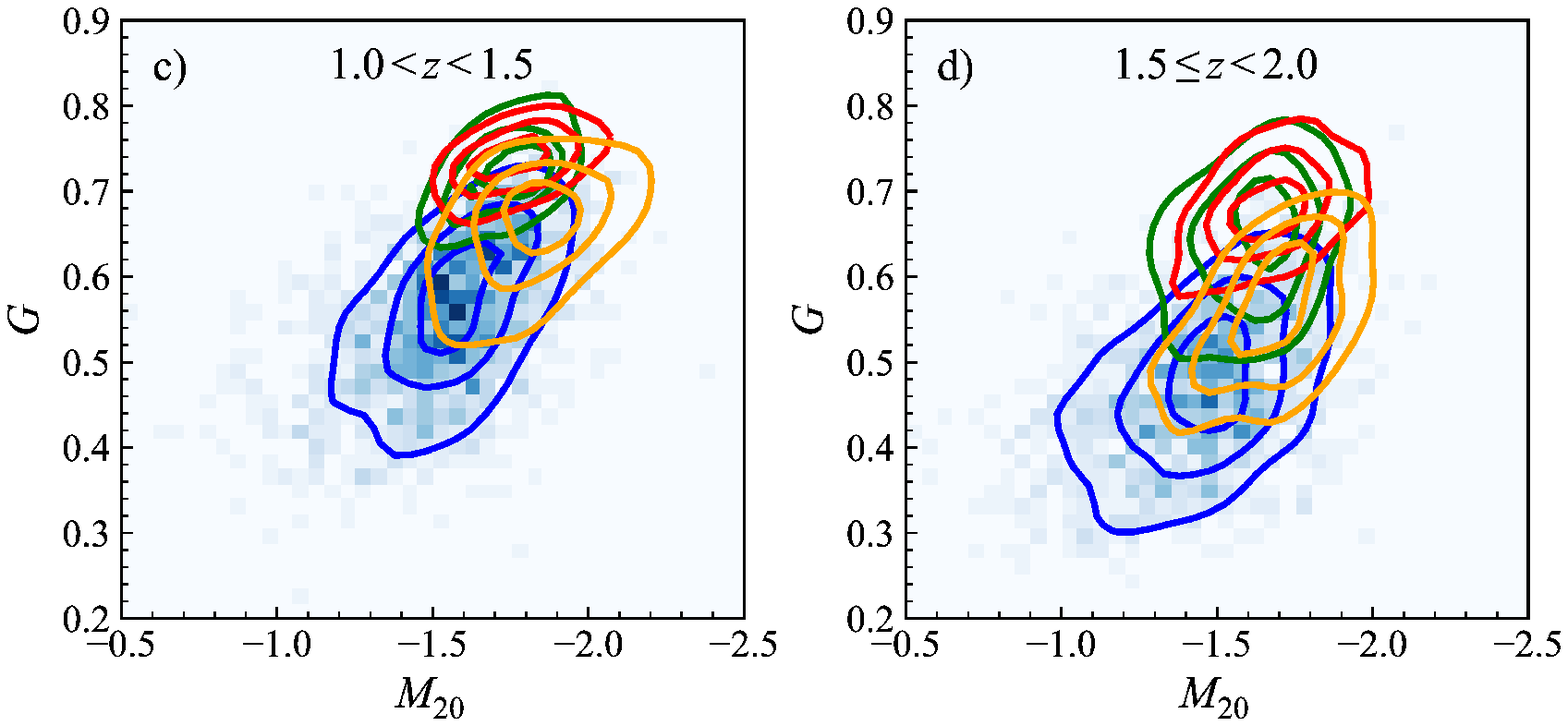}
\caption{Nonparametric morphology of $C$ vs. $M_{20}$ and $G$ vs. $M_{20}$ for four galaxy populations. The overlapped contours include  20\%, 50\%, and 80\% of data points respectively. The colors are the same as those in Figure \ref{fig02}.
}
\label{fig06}
\end{figure*}
Figure \ref{fig06} shows the nonparametric morphologies $C$ and $G$ vs. $M_{20}$ for four galaxy populations at two redshfit bins.
Compact galaxies are substantially offset from eSFGs on the diagrams of nonparametric measurements. The mean values of $C$ and $G$ for cQGs are slightly larger than for cSFGs. Meanwhile, the mean values of $M_{20}$ for cQGs are slightly smaller than those for cSFGs.
The difference in nonparametric morphologies shows that the structures of cQGs are slightly more concentrated and less clumpy than cSFGs, indicating that the quenching processes from cSFGs to cQGs will change their morphologies into more bulge-dominated.

Apart from compact galaxies, eQGs also show prominent offset from eSFGs in the way that
eQGs appear more concentrated and have less substructures than eSFGs.
Several works suggest that the concentrated (or denser) surface density profile is an important condition for quenching (e.g., \citealt{Kauffmann+03, Schiminovich+07, Cheung+12, Fang+13, Lang+14, Barro+17a, Whitaker+17, Gu+18}).
In the early track, compaction is a pre-request for the quenching of compact galaxies, and the rapid quenching process from cSFGs to cQGs also bring about slight bulge growth.
But in the late track, compaction appears to be accompanied by the quenching process due to the sustained buildup of central component. Regardless of the early or late tracks, Figure \ref{fig06} suggests that quenching is associated with compactness.
Interestingly, comparing the nonparametric measurements in two reshift bins,  the buildup of central bulge over cosmic time can be found for all galaxy populations.

\subsection{Role of AGN Feedback on Quenching}
Several works find that the cSFG population has a higher probability holding an AGN. Based on the $Chandra$ 4Ms observation in GOODS-S (\citealt{Xue+11}) and the {\it XMM} 50–100 ks survey in UDS \citep{Ueda+08}, \cite{Barro+13} find the cSFGs at $z > 2$  are 30 times ($\sim 30\%$) more frequently to host X-ray luminous AGNs than the non-compact analogues ($<1\%$).
In GOODS-S/CANDELS, \cite{Barro+14a} find that a higher fraction (47\%) of massive cSFGs at $z \sim 2$ hosts an X-ray AGN, whereas only about 10\% of other massive galaxies host AGNs at the same epoch.  \cite{Kocevski+17} combine $\sim600$ ks catalog from X-UDS survey (PI. G. Hasinger) with other  point-source catalogs publicly drawn from GOODS-S (4 Ms, \citealt{Xue+11}), GOODS-N (2 Ms, \citealt{Xue+16}) and EGS (800 ks, \citealt{Nandra+15}).
They find 39.2\% of cSFGs at $1.4 < z < 3.0$ hosting an X-ray AGN. This fraction is about 3.2 times higher than that of the eSFGs with similar masses.

Suffering from the depth of various X-ray surveys, some low-luminosity and/or highly obscured AGNs may be missed \citep{Ni+19, Yang+19}.
Here, we build a small subsample only in the two Chandra deep fields where the X-ray source catalogs are provided with the exposure of the order of megaseconds:  2 Ms in GOODS-N, \citep{Xue+16} and 7 Ms in GOODS-S \citep{Luo+17}. The main source catalog and supplementary NIR bright catalog provide the well constructed AGN sample in the two fields. AGNs satisfy at least one of the five following criteria. (1) A source with an intrinsic X-ray luminosity; (2) A source with an effective photon index; (3) A source with an X-ray-to-optical flux ratio larger than -1;  (4) A source with excess X-ray emission over the level expected form pure star formation; (5) A source with optical spectroscopic AGN features. The above criteria were described in detail in Xue et al. (2011). To check AGN fraction, we search the AGN counterparts by adopting a matching radius of 1.5 arcsec. Table \ref{tab1} presents the AGN fractions in the fields of GOODS-S and GOODS-N. Statistical uncertainty of AGN fraction is estimated by $\sigma_f = [f_{\rm AGN}(1-f_{\rm AGN})/N_{\rm tot} ]^{1/2}$, assuming the binomial statistics.

Compared with the eSFGs and cQGs at the same epoch, the cSFGs are found to have the {\bf highest} probability to host AGNs. The AGN fraction in cSFGs reaches $\sim 28\%$, significantly higher than those in eSFGs and cQGs. Similarly, a higher incidence of AGN in cSFGs are also found at $2<z<3$ \citep{Barro+13, Barro+14a, Kocevski+17}.
This suggests that the transformation from eSFGs to cSFGs traced by remarkable bulge growth is caused by the centripetal inflow of cold gas,  and the accretion towards central black hole is likely to be driven as well.
In particular,  if the highly dissipational processes such as gas-rich merger and disk instability have triggered the central starburst, it is probable that their central AGN activities are simultaneously triggered.
Shortly after gas is consumed at a high rate of speed, cQGs take shape subsequently. As a result of AGN feedback and rapid gas consumption, the shortage of gas can not support the successive feed of AGN, which leads to a lower AGN fraction in cQGs.

\begin{table}
	\centering
	\caption{ AGN fractions  for four galaxy populations at $1<z<2$ in GOODS-N and GOODS-S  \label{tab1}}
	\begin{tabular}{cccc} 
		\hline
		Population & $\rm N_{GAL}$ & $\rm N_{AGN}$ & $\rm f_{AGN}$ \\
		\hline
eSFGs        & 1102 & 120&10.9$\pm$ 0.9\%  \\
cSFGs        &   47 &  13&27.7$\pm$ 6.5\%  \\
cQGs         &  193 &  17& 8.8$\pm$ 2.0\%  \\
eQGs         &  162 &  19&11.7$\pm$ 2.5\%  \\
Total        & 1504 & 169&11.2$\pm$ 0.8\%\\
		\hline
	\end{tabular}
\end{table}

\subsection{Does Environment Affect Quenching?}
Environment is a crucial factor for galaxy evolution \citep{Muldrew+12, Darvish+15}. In this section, we try to find the environmental effects on compaction and quenching processes by  studying the distributions of localized densities of cQGs and cSFGs. Instead of the separation between the central and the satellite, the environmental density is denoted by the continuous value of local overdensity.

\begin{figure*}
\centering
\includegraphics[scale=0.85]{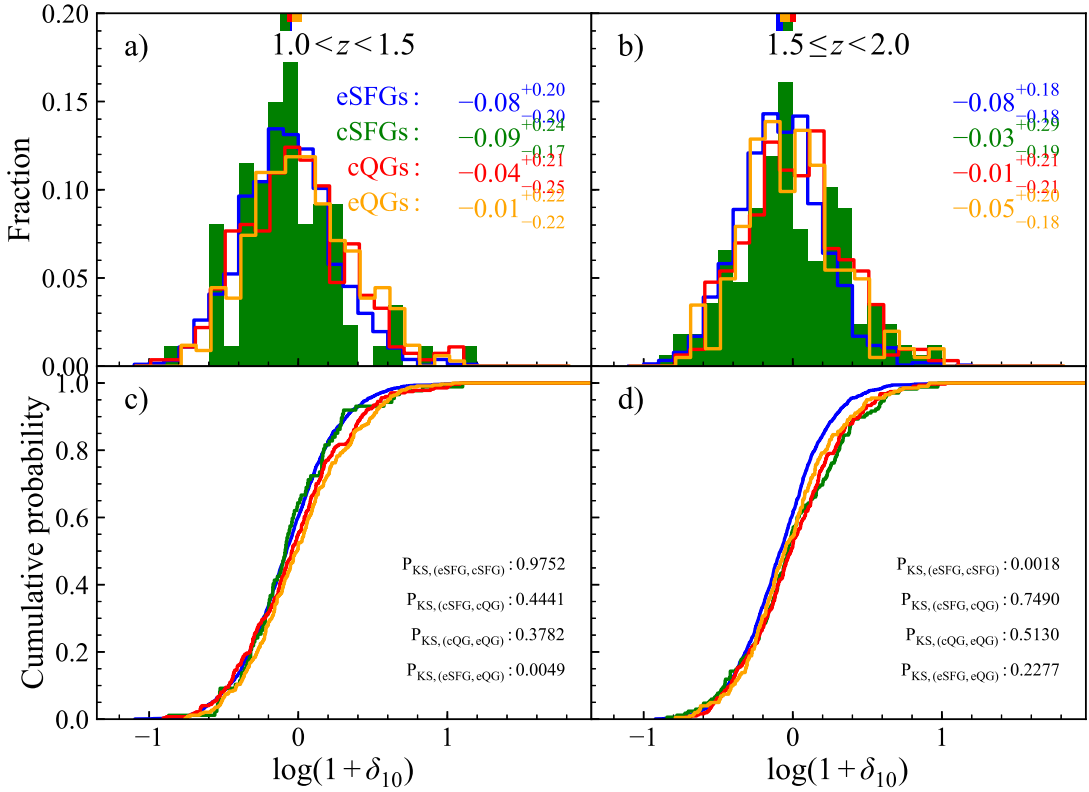}
\caption{Distributions and cumulative probabilities of overdensity in two redshift bins. The eSFGs, cSFGs, cQG, and eQGs are denoted by blue, green, red, and orange histograms, respectively. The median values with uncertainty spanning the 25th to 75th percentiles of each distribution are shown in the corresponding colors. The probabilities of K-S tests between two subsamples are marked in bottom-right panels.
}
\label{fig07}
\end{figure*}

By using low-resolution space-based slitless spectroscopy, the environment of galaxies brighter than $\rm JH_{140} < 24$ mag can be characterized up to $z = 3$ covering $\sim$ 600 arcmin$^2$ in the five 3D--{\it HST}/CANDELS deep fields \citep{Fossati+17}.
Moreover, benefiting from deep and narrow near-IR bands, the FourStar Galaxy Evolution survey ($\sim$ 400 arcmin$^2$; \citealt{Straatman+16}) provides the photometric redshift with high quality to explore the environment for fainter galaxies \citep{Kawinwanichakij+17}. Due to the 3D-{\it HST} project also providing the good photometric redshift over all the $\sim$ 900 arcmin$^2$ region, we mix galaxies with spectroscopic, grism and photometric redshifts for larger sample size. We build a magnitude-limited sample at $z = 1-2$ with $H_{\rm F160W} < 25$ for the measures of environmental density.

The traditional indicator of local environment is defined by the nearest 10 neighboring galaxies, $\Sigma_{10}$, which was firstly proposed by \cite{Dressler+80}.
In this work, this methodology has been improved to measure the local overdensity at high redshifts (Gu et al. 2019, in preparation), by using a Bayesian metric \citep{Ivezic+05, Cowan+08}.
Better than traditional tracer $\Sigma_{10}$, the Bayesian metric considers the distances of all 10 nearest neighbors rather than the distance of only the 10th neighbor, which improves the accuracy of density estimate by a factor of 3.5 in rebuild the probability density distribution (Appendix B2 in \citealt{Ivezic+05}).
For every galaxy in our sample, its local surface density can be estimated by $\Sigma'_{10} = 1/(\Sigma_{i=1}^{10}d^2_{i})$, where $d_i$ is the projected distance (in arcmin) to the $i$th nearest neighbor within a redshift slice of $|\Delta z| < \sigma_{\rm z}(1+z)$. We adopt the factor $\sigma_{\rm z} = 0.02$ as the precision of photometric redshift.
Then, we define the dimensionless overdensity, $1 + \delta'_{10}$, as the environment indicator:
\begin{equation}
1+\delta'_{10} = \frac{\Sigma'_{10}}{\langle \Sigma'_{10}\rangle}=\frac{\Sigma'_{10}}{k'_{10}\Sigma_{\rm surface}},
\end{equation}
where $\Sigma_{\rm surface}$ is the surface number density in the unit of arcmin$^{-2}$ within a given redshift slice. The denominator $\langle \Sigma'_{10}\rangle$ is the typical environmental tracer of Bayesian density when galaxies distribute in the uniform condition at the given $\Sigma_{\rm surface}$.
It can be estimated by the product of $k'_{10}$ and $\Sigma_{\rm surface}$, where $k'_{10}$ is a correction factor which describes the intrinsic linear correlation between $\Sigma_{\rm surface}$ and the Bayesian density $\Sigma'_{10}$. Based on the simulated uniform distributions with a wide range of surface density, $1 < \Sigma_{\rm surface} / {\rm arcmin}^{-2}< 11$, we perform a linear fitting to the correlation between the typical Bayesian density $\langle \Sigma'_{10}\rangle$ and surface density, and achieve $k'_{10}=0.06$.

The overdensity distributions for four galaxy populations in two $z$-bins are shown in the upper panels of Figure \ref{fig07}. The bottom panels present the cumulative distribution functions for the samples of four galaxy populations in order to illuminate the differences in environment.
The Kolmogorov-Smirnov (K-S) tests between the cQGs and eQGs at $1<z<2$ reveal that these two populations have similar $1+\delta'_{10}$ distributions. Environmental densities of eSFGs apparently differ from that of eQGs.
Quiescent galaxies are trend to reside in denser environment, suggesting that the denser environment might have played a role on star formation quenching.

For cSFGs, their preference for environmental density is redshift dependent which implies that the environmental effect along the early track also might be redshift dependent.
At $1.5<z<2$, unlike the eSFGs, the cSFGs  prefer denser environment, very similar as cQGs and eQGs at the same epoch. It hints that merger or strong interaction might have played a role on cSFG formation via compaction process  at $z = 1.5-2$, and  subsequential star formation quenching (from cSFGs to cQGs) is likely to happen very efficiently in denser environment.
However, at lower redshift region $1<z<1.5$, cSFGs prefer to be in lower density environment, similar as eSFGs at the same epoch.
The K-S test with $P_{\rm KS}=0.9752$ between the eSFG and cSFG at $1<z<1.5$ shows that the disk instability of individual galaxies might be the main mechanism of cSFG formation at $1<z<1.5$.
There is no difference in environment between the cSFGs and cQGs at $1<z<1.5$.
\cite{Wang+18} also report that the environments of cSFGs and eSFGs with above $10^{9.5} M_\odot$ and redshift of $0.02 < z < 0.05$ are indistinguishable, whereas QGs more likely reside in more massive halos.
If we take the traditional $\Sigma_{10}$ as the environmental density tracer to do the analysis, the main conclusions do not change.

\section{Discussion}
\label{sec:disc}
\subsection{The formation of cQGs}
\label{sec:cQGs}

Although there are some evidence that the formation of cQGs could occur at $z < 1$ (\citealt{NC+18, NC+19}), it is more expected to happen at $z \sim 2 -3$
(\citealt{Barro+13, Barro+14a, vD+15}), where the gas-rich processes are more possible to happen. The highly dissipational processes, violent disk instability and  merger \citep{Dekel+14,Zolotov+15}, are prevalent to explain the formation of compact galaxies, which can provide the fast and adequate feed of gas in the center region of galaxy, trigger the central star formation and accelerate the compaction process. In addition, the extraneous counter-rotating streams (\citealt{Danovich+15}), the recycled gas (\citealt{Elmegreen+14}) and tidal compression (\citealt{Dekel+03}) can also drive gas into the galactic center and promote the formation of compact core.

Once a disc galaxy becomes compact,  star formation activities in central region should have been triggered by the highly dissipational processes. It is probable that the gas infall will also trigger AGN activity. Thus, both subsequential AGN and stellar feedbacks could be responsible for the quenching of compact galaxies through wiping gas out or just heating. In addition,  the gas in galactic disc is likely to be stabilized by gravitation of its dominant bulge, and this mechanism, so called morphological quenching, can also reduce the intensity of star formation (\citealt{Martig+09}).
Based on the parametric and nonparametric measures of morphology, we have studied the structures of various galaxy populations. For both the quenching track from cSFGs to cQGs and the compaction track from eSFGs to cSFGs, clear evidence of bulge growth have been presented in this work.

As discussed above, the early track which is most expected to occur at $z \sim 2 -3$ may also occur at $z \sim 1 - 2$.
We have shown the distributions of cSFGs in the SFR$-M_*$ plane and in the {\it UVJ} diagram, which supports the early-track scenario that cSFGs  are likely to be a bridge between eSFGs and cQGs.
And the cSFGs and cQGs at $1<z<2$ are found to cover a similar range of stellar mass,  and their median stellar masses are nearly the same. It indicates that the quenching timescale of cSFGs is very short.
Otherwise, since the stellar mass would be substantially assembled during the long term quenching process, the quenched cQGs would have a systematically higher mass distribution than their progenitors --- cSFGs.
A decreasing trend in dust content are found along the sequence from eSFGs, cSFGs to cQGs, which points to the scenario of the early track.
Comparison of stellar age distributions between cSFGs and eSFGs confirms the supposition that the cSFGs are short-lived and will be rapidly quenched into the quiescent phase.
The AGN prevalence in cSFGs at $1<z<2$ indicates that the gas-rich processes are needed and AGN feedback might play a important role on successive quick quenching.
The different preferences to environmental densities between eSFGs and cSFGs support the environmental effects on the compaction along the early track.
The cSFGs at higher redshifts ($1.5<z<2$) prefer dense environment, whereas cSFGs are found to prefer low-density regions at $1<z<1.5$.
This suggests that merger or strong interaction might be the main mechanism of compaction at higher redshifts, whereas the disk instability of individual galaxies might play an more important role on the formation of cSFGs at lower redshifts. As these is no difference in environment between cSFGs and cQGs, it hints that the environment has a little impact on the quenching phase along the early track over our redshift range.

\subsection{The Formation of eQGs}
\label{sec:eQGs}

Observation reveals that quiescent galaxies increase their sizes over cosmic time (e.g., \citealt{vdW+14}).
Both the early and late tracks may be responsible for the build-up of eQG population.
The formation of eQGs can be attributed to the two aspects: the size growth of cQGs and the quenching of eSFGs without the need of compaction phase.
It has been supposed that the size growth of cQGs could be caused by  minor dry merger (\citealt{Hopkins+09b, Hopkins+10}),  AGN/stellar feedback (\citealt{Fan+08, Fan+10}), and mass accretion in outskirts (\citealt{Morishita+16}).
Direct quenching of eSFGs without compaction phase arises from some gas poor processes with longer quenching timescales, such as secular process \citep{KK+04}, strangulation \citep{Peng+15}, and morphological quenching \citep{Martig+09}. None of these mechanisms would bring the significant loss of stellar mass.

In Figure \ref{fig04}, the eQGs at $1.5<z<2$ are found to have intermediate median mass comparing with eSFGs and cQGs.
If all eQGs were formed by the size growth of cQGs, eQGs should have systematically larger stellar masses than cQGs (or similar at least).
It means that majority of eQGs might be formed by some gas-poor quenching processes along the late track.
Although the number of cQGs is larger than that of eQGs at $z \sim 1.5-2$, the early track may not responsible for the majority of eQGs.
The mass distribution of eQGs at $1<z<1.5$ may point to a complex origin of eQG population.
Though it would be hard to distinguish the origin of individual eQGs,  it is probable that both the early track and late track might have contributed to the formation of eQGs.
On one hand, eQGs could increase their stellar masses via dry merger of cQGs, which makes these eQGs have comparable stellar masses relative to compact galaxies.
On the other hand, the newly formed eQGs via direct long-term quenching of eSFGs may have systematically higher stellar mass distribution in comparison with eSFGs.
In our redshift range, the late track should play a role on the formation of eQGs.

\section{Summary}
\label{sec:sum}
In this work, we construct a large sample of massive galaxies ($M_*> 10^{10} M_\odot$) at $1 < z < 2$. By adding all five 3D--{\it HST}/CANDELS fields up,  the influence of cosmic variance can be reduced. After dividing massive galaxies at $z = 1-2$ into four galaxy populations (i.e., eSFGs, cSFGs, cQGs, and eQGs), we investigate their star-formation status, stellar population parameters, structural parameters, AGN fractions and environmental densities.

Our main conclusions are as follows.
\begin{enumerate}
\item The early track which is most expected to occur at $z = 2 -3$ can also occur at $z = 1 - 2$.  We find that the overall level of star formation for cSFGs is relatively lower than that for eSFGs. Only a few cSFGs show starburst feature, which indicates a short life for cSFGs which is due to the high rate of  gas consumption. The results are strongly supported by the {\it UVJ} diagrams and similar mass distributions for cQGs and cSFGs.
\item The distributions of dust attenuation and S{\'e}rsic index support that the progenitors of cQGs are cSFGs, and cSFGs are at a transitional phase when eSFGs come to consume the cold gas and to be quenched into cQGs.  This points to the early-track scenario that compaction is a pre-request for the quenching of compact galaxies.
\item Our analysis of parametric and nonparametric morphologies shows that cQGs (eQGs) are more concentrated and have less substructures than cSFGs (eSFGs).  Quenching and compactness should be associated with each other regardless of the early track or the late track.
\item We confirm the AGN prevalence in cSFGs at $1<z<2$, which indicates that the violent gas-rich interactions such as merger and disk instability could drive the structure to be more compact, and trigger both star formation and black hole growth in the central regions.
\item Quiescent galaxies prefer a denser environments, whereas eSFGs are likely to reside in the lower overdensity.  The cSFGs at $1<z<2$ show a clear redshift dependence of their environmental densities.
The cSFGs at $1.5<z<2$ prefer to be in denser environment, similar to the quiescent galaxies, whereas the cSFGs at $1<z<1.5$ are likely to prefer lower density environment, similar to the eSFGs. It suggests that merger or strong interaction might be the main mechanism of compaction at higher redshifts, whereas the disk instability of individual galaxies might play a more important role on the formation of cSFGs at lower redshifts.
\end{enumerate}

\section*{Acknowledgements}
\addcontentsline{toc}{section}{Acknowledgements}

This work is based on observations taken by the 3D--{\it HST} Treasury Program (GO 12177 and 12328) with the NASA/ESA {\it HST}, which is operated by the Association of Universities for Research in Astronomy, Inc., under NASA contract NAS5-26555.
This work is supported by the National Natural Science Foundation of China (Nos. 11673004, 11433005, 11873032) and by the Research Fund for the Doctoral Program of Higher Education of China (No. 20133207110006). G. W. F acknowledges the support from Yunnan young and middle-aged academic and technical leaders reserve talent program (No. 201905C160039), Yunnan ten thousand talent program - young top-notch talent and Yunnan Applied Basic Research Projects (2019FB007).



\end{document}